\documentclass[genTeX,letterpaper]{nrc1}
\usepackage{graphicx}

\volyear{?}{?}
\received{?}
\accepted{?}

\begin{document}


\newcommand{\eg}{\textit{e.g.}}
\newcommand{\ie}{\textit{i.e.}}
\newcommand{\exteriorDeriv}{\mathrm{d}} 
\newcommand{\integ}[2]{\int_{#1} \exteriorDeriv #2\,}
\newcommand{\integN}[3]{\int_{#1} \exteriorDeriv^{#2}#3\,}
\newcommand{\myVect}[1]{{\mathbf{#1}}} 
\newcommand{\electronMass}{m_{\mathrm{e}}}
\newcommand{\seShift}{{\cal E}_{\rm SE}}
\newcommand{\myL}{L}
\newcommand{\gse}{G_{\mathrm{SE}}}
\newcommand{\gseSeven}{G_{\mathrm{SE,7}}}
\newcommand{\fig}[1]{Fig.~\ref{fig:#1}}
\newcommand{\sect}[1]{Sec.~\ref{sec:#1}}
\newcommand{\citeRefTwo}[2]{Ref.~\cite[#1]{#2}}
\newcommand{\citeRef}[1]{Ref.~\cite{#1}}
\newcommand{\citeRefs}[1]{Refs.~\cite{#1}}
\newcommand{\eq}[1]{Eq.~(\ref{eq:#1})} 
\newcommand{\eqPar}[1]{(\ref{eq:#1})}
\newcommand{\addrFreiburg}{Theoretische Quantendynamik, Physikalisches Institut der Universit\"{a}t Freiburg, Hermann-Herder-Stra\ss e 3, 79104 Freiburg im Breisgau, Germany}
\newcommand{\addrHeidelberg}{Max-Planck-Institut f{\"u}r Kernphysik, Saupfercheckweg 1, 69117 Heidelberg, Germany}
\newcommand{\addrParis}{Laboratoire Kastler Brossel, \'Ecole Normale Sup\'erieure et Universit\'{e} Pierre et Marie Curie, Case 74, 4 place Jussieu,
75005 Paris, France}
\newcommand{\addrGaithersburg}{National Institute of Standards and Technology, Mail Stop 8401, Gaithersburg, MD 20899-8401, USA}


\title{Toward high-precision values of the self energy of non-S states
  in hydrogen and hydrogen-like ions}

\author[E.-O. Le~Bigot]{E.-O. Le~Bigot}
\address{\addrParis}
\author[U.D. Jentschura]{U.D. Jentschura}
\address{\addrHeidelberg}
\author{P. Indelicato}
\address{\addrParis}
\author{P.J. Mohr}
\address{\addrGaithersburg}

\shortauthor{Le~Bigot, Jentschura, Indelicato and Mohr}

\maketitle
\begin{abstract}
  The method and status of a study to provide numerical,
  high-precision values of the self-energy level shift in hydrogen and
  hydrogen-like ions is described. Graphs of the self energy in
  hydrogen-like ions with nuclear charge number between 20 and~110 are
  given for a large number of states.  The self-energy is the largest
  contribution of Quantum Electrodynamics (QED) to the energy levels
  of these atomic systems.  These results greatly expand the number of
  levels for which the self energy is known with a controlled and high
  precision.  Applications include the adjustment of the Rydberg
  constant and atomic calculations that take into account QED effects.
\\\\PACS Nos.: 12.20.Ds, 31.30.Jv, 06.20.Jr, 31.15.-p
\end{abstract}
\begin{resume}
  Nous rapportons les derniers d\'eveloppements d'une \'etude
  destin\'ee \`a fournir des valeurs num\'eriques de grande
  pr\'ecision du d\'eplacement de self-\'energie dans l'hydrog\`ene et
  les ions hydrog\'eno\"\i des. Nous pr\'esentons des graphes de la
  self-\'energie dans les ions hydrog\'eno\"\i des de nombre de charge
  compris entre 20 et~110, pour de nombreux niveaux. La self-\'energie
  est la contribution la plus importante de l'\'electrodynamique
  quantique aux niveaux d'\'energie de ces syst\`emes atomiques. Les
  r\'esultats pr\'esent\'es \'etendent largement l'ensemble des
  niveaux atomiques dont on conna\^\i t le d\'eplacement de
  self-\'energie avec une pr\'ecision importante et contr\^ol\'ee. Les
  applications de ce travail incluent l'ajustement de la constante de
  Rydberg et les calculs atomiques qui prennent en compte les effets
  de l'\'electrodynamique quantique.
\end{resume}

\section{Introduction}

This paper presents graphs of new high-precision numerical values of
the self energy shift in hydrogen-like ions, and describes work in
progress on high-precision self energy values in hydrogen.

The self energy is the dominant contribution of quantum
electrodynamics (QED) to the energy levels of these one-electron
atomic systems (see, \eg, Fig.~2 in \citeRef{beier98}).  The total
energy of the orbiting electron can be expressed as the following sum
(see, \eg, \citeRefs{mohr98,eides2001,shabaev2002}):
\begin{eqnarray*}
\mbox{electron energy} &=& \mbox{Dirac (relativistic) energy} +
\mbox{pure recoil} + \mbox{pure finite nuclear size}\\
&& {} +\mbox{pure QED} {} + \mbox{rest},
\end{eqnarray*}
where the term denoted by ``rest'' encompasses other forces (strong
and weak), as well as mixed effects (\eg, the energy shift due to
recoil in a QED process).
The self energy is generally the largest contribution to the ``pure
QED'' term, and in particular to the Lamb shift~\cite{lamb47}.

The self energy is the QED process in which the electron emits a
photon and re-absorbs it, as is depicted in the corresponding Feynman
diagram:
\begin{equation}
\label{eq:seGraph}
\includegraphics{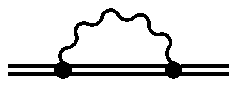},
\end{equation}
where the double line and the wavy line respectively represent the
electron (bound to the nucleus) and a photon.

Calculating QED effects in atoms is of general interest. First,
high-precision spectroscopy results can only be fully explained within
QED; an example is the well-known Lamb shift~\cite{lamb47} and its
calculation by Bethe~\cite{bethe47}. Second, precise atomic QED
calculations, along with corresponding experiments, give some of the
most precise determinations of some fundamental
constants~\cite{mohr2000b}, notably the Rydberg
constant~\cite{biraben2001} and the mass of the
electron~\cite{beier2002b}; the fine structure
constant~$\alpha$~\cite{pachucki2003} and the proton charge
radius~\cite{pohl2001} are also expected to be soon determined with
very high accuracy from atomic studies.  Last, highly-charged
hydrogen-like ions allow for tests of QED in situations where an
electron experiences a strong field. The electric field at the surface
of a uranium nucleus is thus about $2\cdot
10^{21}$~V/m~\cite[p.~230]{mohr98};
this value can for instance be compared to the characteristic field $2
\pi \electronMass c^2/(|e| \lambda_e)$ of
about $10^{18}$~V/m 
that gives a non-negligible probability of spontaneous
electron-positron pair creation in a volume $\lambda_e^3$ in a time
$\lambda_e/c$~\cite[Eq.~(6.41)]{schwinger51}---here, $c$~is the speed
of light, and $\electronMass$, $e$ and $\lambda_e$ respectively denote
the electron mass, charge and Compton wavelength (about
$2\cdot 10^{-12}$~m).
%

\begin{figure}\center\includegraphics[width=0.8\linewidth]{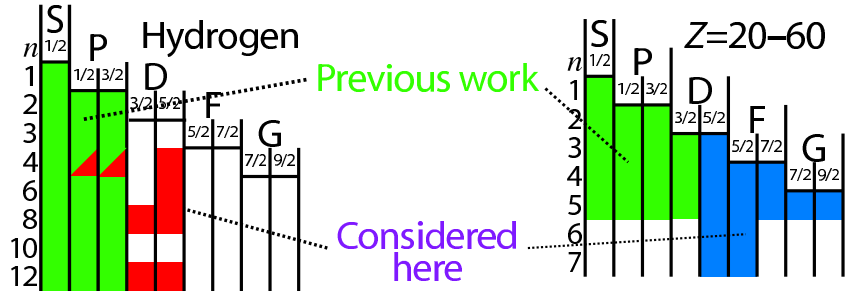}
\caption{\label{fig:knownSEs}%
  Electron states for which a high-precision self-energy shift with
  known uncertainty has been obtained (by either performing direct
  numerical calculations or by using results of such calculations in
  conjunction with perturbation results based on the
  $Z\alpha$-expansion).
$n$~denotes the electron principal quantum number and
  $Z$~is the nuclear charge number of the hydrogen-like ion. Previous
  studies have concentrated on lower-$n$ and lower angular momentum
  states. The results presented here have applications in the
  adjustment of the Rydberg constant~\cite{mohr2004}: the new results
  for hydrogen were obtained for all the non-S levels
  of~\cite[p.~433]{mohr2000b}, \ie, for all the (non-S) levels that
  are implied in the principal spectroscopy experiments contributing
  to the Rydberg constant.  The self energy of the states included in
  the figure can also be used for including QED effects in the energy
  levels of atoms and ions (possibly via extrapolations or
  interpolations).}
\end{figure}

The purpose of this paper is to graphically present new self-energy
shifts in hydrogen-like ions with nuclear charge number larger
than~20. We also report the high-precision evaluation, for many
states, of self energies in hydrogen, and briefly describe the method
we used. All the self energies were calculated for a static, point
nucleus (finite proton mass and non-zero proton radius effects can be
treated as a perturbation).
Hydrogen is a system in which precise numerical calculations that
treat the electron-proton interaction without perturbation are recent
and notoriously
difficult~\cite{jentschura2004,jentschura2001b,jentschura99}---numerical
self-energy calculations were originally designed for highly-charged
hydrogen-like ions~\cite{brown59,brown59a,desiderio71,mohr74,mohr74b}.
The atomic states for which we calculated the self energy are depicted
in \fig{knownSEs}, and extend the set of states considered in previous
works.

The self-energy values considered here were obtained with a high
precision, and a known uncertainty (which contrasts with perturbation
calculations, in which omitted terms can be difficult to
estimate~\cite{mallampalli98}). In hydrogen, we take ``high
precision'' to mean that energy shifts have an uncertainty of about
1~Hz: in fact, spectroscopy experiments are reaching the 1~Hz limit in
the visible domain~\cite{niering2000}. Having high-precision
self-energy shifts available is also useful for obtaining precise
extrapolations (see \sect{hydrogenShift}) and for performing
comparisons between independent calculations.

\section{Notation and outline}

In this paper, the self-energy shifts~$\seShift$ are expressed with
the conventional ``reduced'' self energy~$F$, which is defined by
\begin{equation}
\label{eq:defF}
\seShift(n\myL_j,Z) = \frac{\alpha}{\pi} \frac{(Z\alpha)^4}{n^3}
F(n\myL_j,Z\alpha ) \electronMass c^2,
\end{equation}
where $n\myL_j$~is the usual spectroscopic notation for an
electron of principal quantum number~$n$, orbital angular
momentum~$\myL$ and total angular momentum~$j$; $Z$~is the nuclear
charge number.
Much of the behavior of the self-energy shift~$\seShift$ with $n$
and~$Z$ is captured in this formula: $F$~has the property of not
varying much with these parameters (see, \eg,
\sect{seMiddleHighZ})---the dependence of~$F$ on the angular momenta
$l$ and~$j$ can be estimated from the perturbation expansion of~$F$
(see, \eg, App.~A in \citeRef{mohr2000b}).

In the following, we present graphs of the self-energy
values for middle-$Z$ hydrogen-like ions (\sect{seMiddleHighZ}), and
describe the way self energies were obtained in hydrogen
(\sect{hydrogenShift}).

\section{Self energy in middle-$Z$ ions}
\label{sec:seMiddleHighZ}

\begin{figure}\center\includegraphics[width=0.8\linewidth]{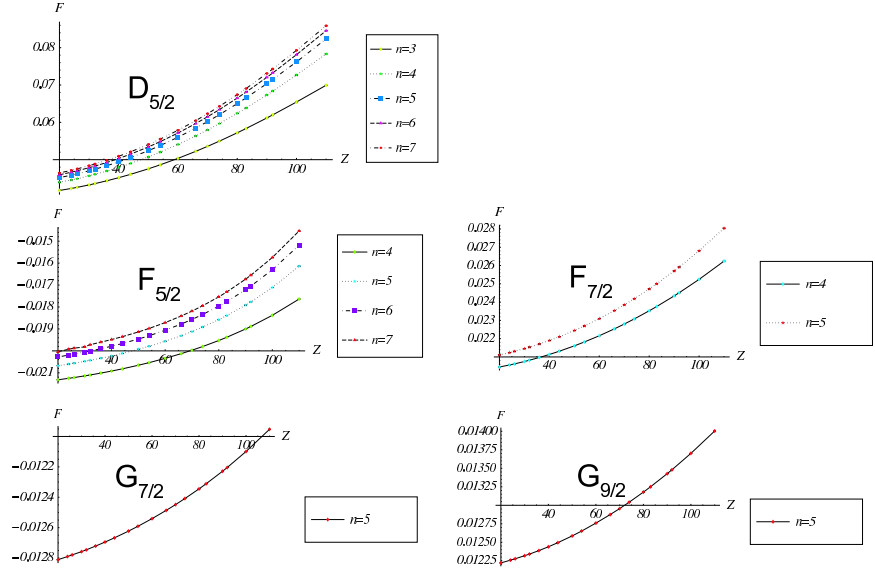}
\caption{\label{fig:seValues}%
  Values of the reduced self energy~$F$, for various D, F and G states
  of hydrogen-like ions, as a function of the nuclear charge
  number~$Z$ of the nucleus and of the principal quantum number~$n$ of
  the electron. The values for $Z < 60$ are new, while higher-$Z$
  results can be found in~\cite{lebigot2001b}. The weak dependence
  of~$F$ on $n$ and~$Z$ indicates that the self energy
  shift~$\seShift$ scales essentially as $Z^4/n^3$, as can be seen
  from \eq{defF}.}
\end{figure}

This section presents graphs of self-energy values for hydrogen-like
atoms of nuclear charge number~$Z$ comprised between 20 and~60.
Figure~\ref{fig:knownSEs} shows the states for which self-energy
results are presented.  The nucleus is approximated as a point-like
charge of infinite mass.  The main result of this part is displayed in
\fig{seValues}, under the form of the reduced self energy~$F$ defined
in \eq{defF}. In the graphs of \fig{seValues}, each curve point took
about two days of calculation time on a modern computer. The typical
relative precision of these self-energy values is $10^{-5}$ (numerical
values will be given elsewhere). As can be seen in the graphs, the
reduced self energy of D, F and G states does not vary much with $n$
and~$Z$ (changes of the order of 10~\%). This implies that
extrapolations of the self energy to states with other principal
quantum numbers than those considered in \fig{seValues} can be done to
a good precision. We expect such a smooth behavior of~$F$ with $n$
and~$Z$ to generally hold for electrons of higher angular momentum.

Calculating the self energy shift consists essentially in evaluating a
multi-dimensional integral (see, \eg, \citeRef{mohr98}):
\begin{equation}
\label{eq:integralToBeCalculated}
\seShift(n\myL_j,Z) =
\integ{\ }{z} \integN{\ }{3}{x_2} \integN{\ }{3}{x_1}
\phi^{\dagger}(\myVect{x}_2)
{\mathcal G}(\myVect{x}_2, \myVect{x}_1, z)
\phi(\myVect{x}_1),
\end{equation}
where $\phi$~is the Dirac wave-function of the state for which the
shift is calculated, and where $\mathcal G$~contains the effect of
both the photon and the electron propagating between the two
interaction points in \eqPar{seGraph}. In \eq{integralToBeCalculated},
the variable~$z$ represents the virtual energy of the photon, and the
positions $\myVect{x}_1$ and~$\myVect{x}_2$ are integrated over all
the possible locations of the two interaction points
in~\eqPar{seGraph}.

In order to compute~\eqPar{integralToBeCalculated}, we used a method
developed by one of us (P.J.M.~\cite{mohr74,mohr74b}), along with the
corresponding code. Contrary to methods introduced
before~\cite{brown59,desiderio71}, it allows for precise calculations
of the electron self energy over a large range of nuclei
($Z=5$--$110$~\cite{mohr92b}); this was made possible through the use of
analytic expressions of the Dirac Green's function, and of an
integration order in~\eqPar{integralToBeCalculated} that guarantees
good convergence properties in numerical calculations.  The code was
incrementally improved along two directions: a more
efficient handling of renormalization~\cite{indelicato98}, and the
possibility of calculating the self energy of states with ``large''
total angular momentum $j>3/2$~\cite{lebigot2001b}.  The calculation
of~\eqPar{integralToBeCalculated} requires high-precision evaluations
of integrals and of hypergeometric functions (and in particular
Whittaker functions).

\begin{figure}\center\includegraphics[width=0.4\linewidth]{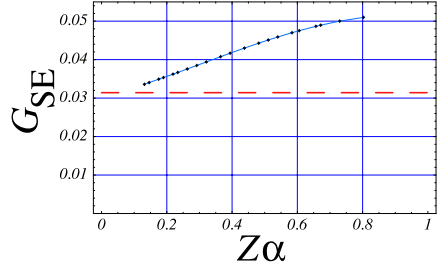}
\caption{\label{fig:GSE_4d5}
  Plot of the self-energy remainder $\gse(Z\alpha)$ defined in
  \eq{Fperturbation}, and calculated with the values displayed in
  \fig{seValues}, for the 4D$_{5/2}$ level. The dashed line represents
  an independent calculation of $\lim_{Z\to 0} \gse(Z\alpha)$. The
  curve is compatible with a convergence of the numerical values of
  $\gse$ toward the correct limit. Similar checks of the new
  self-energy values were performed for all the other levels
  considered in \fig{seValues}.}
\end{figure}

The self-energy values in \fig{seValues} were checked against
perturbation theory results. In fact, the reduced self energy of non-S
states can be expanded as (see, \eg, App.~A in \citeRef{mohr2000b}):
\begin{equation}
\label{eq:Fperturbation}
F(Z\alpha) = A_{40} + (Z\alpha)^2 \{ A_{61} \ln[(Z\alpha)^{-2}]
                                   + \gse(Z\alpha) \},
\end{equation}
where the dependence of the terms on the electron state $n\myL_j$ is
implicit; $A_{40}$~is a coefficient which has been calculated for many
states; $A_{61}$ is known analytically; $\gse(Z\alpha)$ is a remainder
whose limit as $Z\to 0$ has recently been calculated for all the
states considered here~\cite{jentschura2003b,lebigot2003c}.  In order
to check our calculations of~$F$, we made sure that the numerical
values of $\gse(Z\alpha)$ obtained from~$F$ and \eq{Fperturbation}
were compatible with a convergence to the correct limit as $Z\alpha
\to 0$. This procedure also implicitly checks the compatibility of the
new numerical values of~$F$ with the values of $A_{40}$ and~$A_{61}$.
An example of the curves that we plotted for verification purposes is
given in \fig{GSE_4d5}.

%
The range of nuclear charges under consideration ($20\leq Z < 60$)
extends beyond the range of results of \citeRef{lebigot2001b}.
The orbital angular momentum of an $n=5$ electron in the ground state
of atoms can reach $l = 3$ (f~states), and that of an $n=6$ electron,
$l =2$ (d~states); the self energy was thus calculated for both 5f and
6f~states (see \fig{seValues}), so as to provide values that could be
used in atomic calculations that incorporate QED corrections (see also
\citeRef{indelicato98a}).
The self energy was also calculated for higher-$n$ d and f~electrons,
so as to cover some excited atomic states.
%
Moreover, spectroscopy
experiments with middle-charge hydrogen-like ions have been, and may
be performed in the future: the self-energy values in \fig{seValues},
which represent the dominant QED shift, might be useful in
interpreting coming experiments. Finally, these results, which do not
treat the electron-nucleus interaction with perturbation theory, allow
for checks of corresponding perturbation results~\cite{lebigot2003c}.

\section{Self energy in hydrogen}
\label{sec:hydrogenShift}

This section describes how we obtained self-energy shifts in hydrogen
through interpolations and extrapolations, to a high precision (better
than 2~Hz) for the hydrogen states included in \fig{knownSEs}.
As for the middle-$Z$ results, only the leading self-energy
contribution~$\seShift$, obtained for an infinite-mass, point-like nucleus, is
discussed here.
With the extrapolation method described below, we obtained for instance
the self energy shift value 
\begin{equation}
\label{eq:F4P1_exact}
\seShift(4{\mathrm P}_{1/2}, Z=1) = -1\,404.240(2) \mbox{ kHz}.
\end{equation}
This value is consistent with previous estimates of the 4P$_{1/2}$
self energy~(see ~\citeRef{jentschura97} and
\citeRefTwo{p.~468}{mohr2000b}).  Actual self energy values for the
states included in \fig{knownSEs} will be given elsewhere.

Value~\eqPar{F4P1_exact} can be
compared to perturbation calculations. Thus, for non-S states, the
reduced self energy in \eq{defF} is expanded as
\begin{equation}
\label{eq:Fperturbation3coefs}
F(Z\alpha) = A_{40} + (Z\alpha)^2 \{ A_{61} \ln[(Z\alpha)^{-2}]
                                   + A_{60} \}
                                 + (Z\alpha)^3 \gseSeven(Z\alpha)
,
\end{equation}
where $A_{60}$ is a coefficient and $\gseSeven(Z\alpha)$ a function
which converges as $Z\alpha\to 0$~\cite{karshenboim94,karshenboim95}.
No other coefficient than the three present
in~\eqPar{Fperturbation3coefs} is currently known; references about
the known coefficients can for instance be found
in~\cite{lebigot2003c}.  It is instructive to truncate
expansion~\eqPar{Fperturbation3coefs} to the first two terms, and
estimate the self energy of a 4P$_{1/2}$ electron as
\begin{equation}
\label{eq:F4P1_2coefs}
\seShift(4{\mathrm P}_{1/2}, Z=1) = -1\,403.5(7?) \mbox{ kHz},
\end{equation}
where the uncertainty is defined by letting $A_{60} +
\alpha\gseSeven(\alpha)= 0(1?)$, and where question marks indicate
that the uncertainty is only an estimate. 
The crude estimate used here is intended to illustrate the usefulness
of exact numerical calculations (as opposed to pure perturbation
calculations) in situations where not much is known about the
perturbation remainder. The numerical examples we use here are given
for demonstration purposes only (in fact, in the particular case of
the 4P$_{1/2}$ level, the remainder has been previously evaluated to
good accuracy~\cite{jentschura97}).
Doing the same with the
first three terms in \eqPar{Fperturbation3coefs}---\ie\ with all the
terms calculated with perturbation theory---yields
\begin{equation}
\label{eq:F4P1_3coefs}
\seShift(4{\mathrm P}_{1/2}, Z=1) = -1\,404.260(6?) \mbox{ kHz},
\end{equation}
where the value $\gseSeven(\alpha) = 0(1?)$ was assigned. By comparing
\eqPar{F4P1_2coefs} and~\eqPar{F4P1_3coefs} with the exact
result~\eqPar{F4P1_exact}, we see that the exact result is
more precise than the perturbation estimates. It also has the
advantage of providing an error estimate which is more reliable than
those of the perturbation results.

\begin{figure}\center\includegraphics[width=0.7\linewidth]{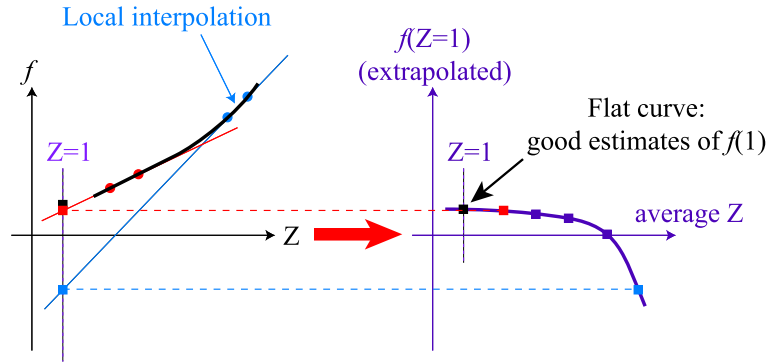}
\caption{\label{fig:localFits}Principle of the interpolation/extrapolation method~\cite{mohr75,lebigot2003c} used in obtaining self energies in hydrogen ($Z=1$) from middle-$Z$ self energies.
  In order to estimate the limit of a function~$f(Z)$ as $Z\to 1$, it
  is possible to plot its curve and visually estimate the limit along
  with an uncertainty (left-hand graph). A more precise estimate can
  be obtained by performing a linear interpolation between two
  successive points of the curve, and plotting the ordinate of the
  line at its intersection with the $Z=1$ axis, as a function of the
  average~$Z$ of the two points (right-hand graph). This process can
  be generalized to polynomial interpolations of each group of three
  (or more) successive data points of~$f(Z)$: each interpolation has a
  value at $Z=1$, which can be plotted as in the right-hand graph.
  Higher-order interpolations can yield more and more precise
  estimates of $\lim_{Z\to 1} f(Z)$.}
\end{figure}

Our high-precision self-energy values were obtained by using
middle-$Z$ numerical self energies and performing interpolations and
extrapolations to hydrogen, \ie, to $Z=1$.
Similar methods were used in the past, and yielded very accurate
self-energy values (see
\citeRefs{jentschura97,karshenboim94,karshenboim95,mohr75,jentschura96,mohr91,mohr95}
and \citeRefTwo{p.~468}{mohr2000b}).
With the recent application~\cite{jentschura99} of convergence
acceleration techniques to the calculation method we use, a direct
numerical computation of self-energy shifts in hydrogen could have
been possible for any~\cite{lebigot2001b} of the states of hydrogen
depicted in \fig{knownSEs}. However, the computation of the
self-energy shift of one $n\myL_j$ level takes a few weeks, on a
modern processor, and a simpler method can yield values with an
accuracy better than the experimental precision.
We thus performed extrapolations of middle-$Z$ self-energy shifts or
interpolations between perturbation and exact results
(middle-$Z$ self energies were found in previously published works, or
taken from the results presented in \sect{seMiddleHighZ}).
In order to obtain high-precision shifts, two
strategies were used.  First, we employed the general
interpolation/extrapolation method described in the appendix of
\citeRef{lebigot2003c}, and which was first sketched
in~\citeRef{mohr75}. The principle of this method is given in
\fig{localFits}.  Second, the extrapolated (or interpolated) quantity
was not~$F$ itself, but ``remainders'' of~$F$: we extrapolated (or
interpolated) to $Z=1$ both $ \gseSeven(Z\alpha)$ and $ A_{60} +
(Z\alpha) \gseSeven(Z\alpha) $---we used \eq{Fperturbation3coefs} as a
magnifying glass that can show small variations in~$F$. Since no
middle-$Z$ self energies are available for the $n=8$ and $n=12$ states
included in \fig{knownSEs}, we also performed extrapolations to these
principal quantum numbers; such extrapolations can be performed with
good accuracy because self-energy values do not vary much with~$n$, as
is depicted in \fig{seValues}. More details on the interpolation and
extrapolation procedures that yielded the self energy for the other
hydrogen states included in \fig{knownSEs} will be given elsewhere.

The hydrogen states for which we calculated the self energy (see
\fig{knownSEs}) represent all the \mbox{non-S} states used in the
CODATA98 adjustment of the fundamental
constants~\cite[p.~433]{mohr2000b}---these states are all involved in some
high-precision spectroscopy experiment. The self-energy values were
used in the latest CODATA adjustment~\cite{mohr2004}.  The precisions
of our self energy values are below~2~Hz, which is smaller than the
uncertainty in the best experimental frequency measurement ever
achieved in the optical domain (about 50~Hz~\cite{niering2000}).

\section{Conclusion}

We have discussed high-precision numerical evaluations of the self
energy~\eqPar{seGraph}, which is the dominant QED contribution to the
energy levels of hydrogen and of hydrogen-like ions. The atomic levels
which were considered are depicted in \fig{knownSEs}.
They greatly extend the set of states for which the self energy is
known with a high precision and a controlled uncertainty.
The results that are reported here for hydrogen were used in the
latest CODATA adjustment of the fundamental constants~\cite{mohr2004}.
The middle-$Z$ results can find application in atomic calculations
that include QED contributions. The high precision of all the results
can also allow for comparisons with results obtained through
independent methods, or for extrapolations (or interpolations) to
other states or nuclear charge numbers than those considered in this
paper.

The self-energy values in hydrogen considered here were obtained by
taking advantage of the availability of both perturbation and exact
numerical results (which permitted interpolations and precision
extrapolations).  Nonetheless, it is desirable to perform a direct
numerical evaluation of these energy shifts, and check their
consistency with the values presented here.

\section*{Acknowledgments}

E.O.L acknowledges support by NIST\@.  We would like to thank the CINES
(Montpellier, France) and the IDRIS (Orsay, France) for grants of time
on parallel computers. The \textit{Laboratoire Kastler Brossel} is the
\textit{Unit\'e Mixte de Recherche}~8552 of the CNRS\@.


\newcommand{\noopsort}[1]{}


\end{document}